\documentstyle[aps,epsf,eqsecnum,pre]{revtex}

\begin{document}
\draft

\def\g{\gamma}
\def\b{\beta}
\def\a{\alpha}
\def\l{\Lambda}
\def\lan{\lambda_n}
\def\bk{\frac{\b^2}{k}}
\def\NI{\noindent}
\def\qa{{\hat q}_a}
\def\qb{{\hat q}_b}
\def\z{{\overline{z}}}
\def\k{{\overline{k}}}
\def\half{{1\over2}}
\def\kR{k_{\rm R}}
\def\kI{k_{\rm I}}
\def\C{{\rm\bf C}}
\def\O{{\cal O}}
\def\phi{\varphi}
\def\kj{\kappa_j}
\def\mj{\mu_j}
\def\lj{\lambda_j}
\def\und#1{$\underline{\smash{\hbox{#1}}}$}

\title{The Solution of the Modified Helmholtz Equation in a Triangular Domain
       and an Application to Diffusion-Limited Coalescence}

\author{Daniel ben-Avraham$^1$\footnote{{\bf e-mail:} benavraham@clarkson.edu}
 and Athanassios S. Fokas$^2$\footnote{Permanent address:
Department of Mathematics, Imperial College, London SW7 2BZ, UK.  
{\bf e-mail:} a.fokas@ic.ac.uk}}
\address{$^1$ Physics Department, and Clarkson Institute for Statistical
Physics (CISP),\\
Clarkson University, Potsdam, NY 13699-5820\\
$^2$ Institute for Nonlinear Studies, Clarkson University, Potsdam, NY
13699-5805}
\maketitle
\begin{abstract}
A new transform method for solving boundary value problems for linear and
integrable nonlinear PDEs recently introduced in the literature is used here
to obtain the solution of the modified Helmholtz equation
$q_{xx}(x,y)+q_{yy}(x,y)-4\b^2q(x,y)=0$ in the triangular domain
$0\leq x\leq L-y \leq L$, with mixed boundary conditions.  This
solution is applied to the problem of
diffusion-limited coalescence, $A+A\rightleftharpoons A$, in the segment
$(-L/2,L/2)$, with traps at the edges.
\end{abstract}


\section{introduction}

A new method for solving boundary value problems for linear and for
integrable nonlinear PDEs has been introduced recently~\cite{fokas00}.  Here
we apply this method to the equation
\begin{equation}
\label{E.eq}
E_{xx}+E_{yy}+\g(-E_x+E_y)=0\;,  
\end{equation}
in the triangular domain $-L/2\leq x\leq y\leq L/2$, where $E(x,y)$ is a
scalar function and
$\gamma$ is a positive constant.  A solution of Eq.~(\ref{E.eq}) in the
semi-infinite  wedge $0\leq x\leq y$ has been
presented in~\cite{dba99}.  Using the substitution
$E(x,y)=1-e^{-{\g\over2}(y-x)}q(x,y)$, Eq.~(\ref{E.eq}) becomes the
modified Helmholtz equation
\begin{equation}
\label{helmholtz}
q_{xx}+q_{yy}-4\b^2q=0\;,
  \qquad \b={\gamma\over\sqrt{8}}\;.  
\end{equation}
Eq.~(\ref{E.eq}) with
$\gamma=v/D$ represents the steady state of the diffusion-limited reaction
$A+A\rightleftharpoons A$ on the line, where the $A$-particles diffuse with
diffusion constant $D$, they merge immediately upon encounter, and split into
two particles (the back reaction) at rate
$v$~\cite{coalescence,doering92,dba.book}. 
$E(x,y)$ represents the probability that the interval
$(x,y)$ is empty.  The concentration profile of the particles is related to
$E(x,y)$ through $c(x)=-E_y(x,x)$.  Suppose that we limit ourselves to 
the segment $-L/2\leq x\leq L/2$, then the domain of Eq.~(\ref{E.eq}) is
$-L/2\leq x\leq y\leq L/2$. The forward reaction is described by the boundary
condition (BC)
$E(x,x)=1$. If there are perfect traps at the edges, $x=\pm L/2$, one gets the
BCs $E_x(-L/2,y)=0$, $E_y(x,L/2)=0$. These BCs transform into the following
BCs for Eq.~(\ref{helmholtz}):
\begin{eqnarray*}
q(x,x)&=&0\;,\qquad -\case{L}{2}\leq x\leq\case{L}{2}\;,\\
\case{\g}{2}q(-\case{L}{2},y)+q_x(-\case{L}{2},y)&=&0\;,
  \qquad -\case{L}{2}\leq y\leq\case{L}{2}\;,\\
-\case{\g}{2}q(x,\case{L}{2})+q_x(x,\case{L}{2})&=&0\;,
  \qquad -\case{L}{2}\leq x\leq\case{L}{2}\;.
\end{eqnarray*}
We rotate and translate the $(x,y)$-axes, with the mapping
$(x,y)\mapsto(-y+L/2,x+L/2)$. Eq.~(\ref{helmholtz}) remains
invariant, but the domain is now $0\leq x\leq L-y \leq L$ --- the isosceles
right triangle with vertices at $(0,0)$, $(0,L)$, $(L,0)$ --- and the BCs
become
\begin{mathletters}
\label{bcs}
\begin{eqnarray}
q(x,L-x)&=&0\;,\qquad 0\leq x\leq L\;,\\
\case{\g}{2}q(x,0)+q_y(x,0)&=&0\;,\qquad 0\leq x\leq L\;,\\
\case{\g}{2}q(0,y)+q_x(0,y)&=&0\;,\qquad 0\leq y\leq L\;.
\end{eqnarray}
\end{mathletters}
For the sake of generality, instead of the BC~(\ref{bcs}b), (\ref{bcs}c) we
consider
\begin{eqnarray}
\case{\g}{2}q(x,0)+q_y(x,0) &=& f(x)\;,\qquad 0\leq x\leq L\;, 
  \eqnum{\ref{bcs}b'} \\
\case{\g}{2}q(0,y)+q_x(0,y) &=& f(y)\;,\qquad 0\leq y\leq L\;, 
  \eqnum{\ref{bcs}c'}
\end{eqnarray}
where $f(\cdot)$ is an arbitrary smooth function. 

We will show that: (a)~Eq.~(\ref{helmholtz}) with the BCs~(\ref{bcs}a),
(\ref{bcs}b'), (\ref{bcs}c') has  a unique solution that can be expressed in
closed form.  (b)~Eq.~(\ref{helmholtz}) with the homogeneous
BCs~(\ref{bcs}a),  (\ref{bcs}b), (\ref{bcs}c) has only the trivial solution
$q(x,y)=0$, {\it i.e.}, the only steady state of the process
$A+A\rightleftharpoons A$, in a segment demarcated by traps, is the vacuum ---
when there are no particles left --- regardless of the magnitude of $v$, the
rate of the back reaction $A\to A+A$.  (c)~For large back reaction rates,
$\g L\gg1$, the characteristic relaxation time to the empty,
absorbing state grows exponentially as $(D/2v^2)e^{vL/2D}$.

Let $z=x+iy$, let a bar denote the complex conjugate ($\z=x-iy$), and let
$z_j$ denote the corners of the domain
$0\leq x\leq L-y \leq L$ (see Figure~1);
\begin{equation}
\label{3}
z_1=L,\qquad z_2=0,\qquad z_3=iL\;. 
\end{equation}

\section{the analysis of the inhomogeneous problem}
It is shown in~\cite{fokas01} that the general solution of the modified
Helmholtz equation in the above domain can be represented as
\begin{equation}
\label{2.1}
q(x,y)=\frac{1}{2\pi i}\sum_{j=1}^3\int_{\ell_j}
   e^{ikz-i\bk\z}\rho_j(k)\frac{dk}{k}\;, 
\qquad 0\leq x\leq L-y\leq L\;, 
\end{equation}
where $\ell_1$, $\ell_2$, $\ell_3$, are the rays on the complex $k$-plane
defined by $\arg k=0$, $\case{\pi}{2}$, $\case{5\pi}{4}$, and oriented from
zero to infinity (see Figure~2), while the functions $\rho_j(k)$ are defined
by
\begin{equation}
\label{2.2}
\rho_j(k)=\int_{z_{j+1}}^{z_j}e^{-ikz+i\bk\z}[\case1/2(q_x-iq_y)\,dz
   +i\case{\b^2}{k} q\,d\z]\;,
\qquad k\in\C\;,\qquad j=1,2,3\;, \qquad z_4=z_1\;. 
\end{equation}
Using the boundary conditions~(\ref{bcs}) to simplify the expressions for
$\rho_j(k)$, we find the following:
\begin{mathletters}
\label{2.3}
\begin{eqnarray}
\rho_1(k) &=& -\case1/2 q(0,0)+i\a(k)\psi_1(-ik)-iF(-ik)\;, 
  \quad k\in \C\;, \\
\rho_2(k) &=& \case1/2 q(0,0) +i\a(-ik)\psi_2(k)-iF(k)\;, 
  \quad\qquad k\in \C\;, \\
\rho_3(k) &=& iE(k)\psi_3(-ke^{i\frac{\pi}{4}})\;, 
  \quad\qquad\qquad\qquad\qquad\quad k\in \C\;,
\end{eqnarray}
\end{mathletters}
where
\begin{equation}
\label{2.4}
\a(k)=\case1/2(\case{\b^2}{k}+k+\case{\g}{2}), \qquad
  E(k)=e^{(k+\bk)L}\;,\qquad F(k)=\frac{1}{2}\int_0^Le^{(k+\bk)y}f(y)\,dy\;,  
\end{equation}
and the unknown functions $\psi_1$, $\psi_2$, $\psi_3$ are defined by
\begin{equation}
\label{2.5}
\psi_1(k) = \int_0^Le^{(k+\bk)x}q(x,0)\,dx\;, \quad
\psi_2(k) = \int_0^Le^{(k+\bk)y}q(0,y)\,dy\;, \quad
\psi_3(k) = \int_0^{\sqrt{2}L}e^{(k+\bk)s}
                      q_s(\case{s}{\sqrt{2}},L-\case{s}{\sqrt{2}})\,ds\;.  
\end{equation}
Indeed, for the derivation of~(\ref{2.3}a) we use $z=x$, and we note that
the boundary condition~(\ref{bcs}b') implies
\[
\case1/2(q_x(x,0)-iq_y(x,0))+i\case{\b^2}{k}q(x,0)
   =\case1/2q_x(x,0)+i(\case{\b^2}{k}+\case{\g}{4})q(x,0)
     -\case{i}{2}f(x)\;;
\]
integrating by parts the terms involving $q_x$ we find~(\ref{2.3}a).  The
derivation of~(\ref{2.3}b) is similar, where we use the
condition~(\ref{bcs}c').  For the derivation of~(\ref{2.3}c) we use
$z=iL+x-ix$, and we note that the boundary condition $q(x,L-x)=0$ implies
$q_x(x,L-x)-q_y(x,L-x)=0$.

In order to simplify the analysis, we have assumed that the {\em same}
function $f$ appears in the BCs~(\ref{bcs}b') and (\ref{bcs}c').  This
implies that the PDE~(\ref{helmholtz}), the triangular domain, and the
BCs~(\ref{bcs}a), (\ref{bcs}b'), (\ref{bcs}c') are invariant under the
reflection $x\leftrightarrow y$, thus $q(x,y)=q(y,x)$.  Hence,
$\psi_1(k)=\psi_2(k)$.

We introduce the following notations:
\begin{equation}
\label{2.6}
\psi_1(k)=\psi_2(k)=\phi(k)\;,\qquad\psi_3(-ke^{-i\pi/4})=\psi(-)\;,
  \qquad\psi_3(-ke^{i\pi/4})=\psi(+)\;,\qquad e(k,z,\z)=e^{ikz-i\bk\z}\;.
\end{equation}

\subsection{The Analysis of the Global Relation}
Eqs.~(\ref{2.3}) express $\rho_j(k)$ in terms of the unknown functions
$\phi(-ik)$, $\phi(k)$, and $\psi(+)$.  These functions satisfy the {\em
global condition}: $\sum_{j=1}^3\rho_j(k)=0$~\cite{fokas01}.  This equation,
and its complex conjugate, are
\begin{eqnarray}
\a(k)\phi(-ik)+\a(-ik)\phi(k)+E(k)\psi(+)&=&F(k)+F(-ik)\;,
  \qquad k\in\C\;, \label{2.7}\\ 
\a(k)\phi(ik)+\a(ik)\phi(k)+E(k)\psi(-)&=&F(k)+F(ik)\;,
  \qquad k\in\C\;. \label{2.8}
\end{eqnarray}

Following~\cite{fokas01} we supplement these equations with the equations
obtained from Eqs.~(\ref{2.7}), (\ref{2.8}) by using the
transformations in the complex $k$-plane which leave invariant the {\em
pairs} $\{\phi(-ik),\phi(ik)\}$ and $\{\psi(+),\psi(-)\}$.  The first pair is
invariant under $k\mapsto-k$, and the second pair is invariant under
$\{k\mapsto-ik,k\mapsto ik\}$.  Using the latter transformations,
Eqs.~(\ref{2.7}) and (\ref{2.8}) yield
\begin{eqnarray}
\a(-ik)\phi(-k)+\a(-k)\phi(-ik)+E(-ik)\psi(-)&=&F(-ik)+F(-k)\;,
  \qquad k\in\C\;, \label{2.9}\\ 
\a(ik)\phi(-k)+\a(-k)\phi(ik)+E(ik)\psi(+)&=&F(ik)+F(-k)\;,
  \qquad k\in\C\;. \label{2.10}
\end{eqnarray}
Eqs.~(\ref{2.7})--(\ref{2.10}) are invariant under $k\mapsto-k$, thus
we do not obtain any additional equations using this transformation. 
Eqs.~(\ref{2.7})--(\ref{2.10}) are the basic equations needed for the
determination of the unknown functions $\phi(k)$, $\phi(-ik)$, $\psi(+)$. 
The analysis of the basic equations leads to a matrix Riemann-Hilbert
problem.  However, in what follows we will show that this problem can be
bypassed, and that $q(x,y)$ can be obtained using only {\em algebraic
manipulations} of the basic equations.

Eqs.~(\ref{2.7})--(\ref{2.10}) imply that $\phi(-ik)$, $\phi(k)$, 
$\psi(+)$ can be expressed in terms of $\phi(ik)$ and $\psi(-)$:
\begin{mathletters}
\label{2.11}
\begin{eqnarray}
&&\phi(-ik)=A(-ik)\phi(ik)+\frac{E(k)}{\a(-k)\Delta(k)}[A(-ik)^2E(ik)-E(-ik)]
  \psi(-)+G_1(k)\;,\\
&&\phi(k)=-\frac{\a(k)}{\a(ik)}\phi(ik)-\frac{E(k)}{\a(ik)}\psi(-)
  +G_2(k)\;,\\
&&\psi(+)=\frac{A(-ik)E(k)+A(k)E(-ik)}{\Delta(k)}\psi(-)+G_3(k)\;,
\end{eqnarray}
\end{mathletters}
where
\begin{equation}
\label{2.12}
A(k)=\frac{\a(k)}{\a(-k)}\;, \qquad \Delta(k)=E(k)+A(k)A(-ik)E(ik)\;,
\end{equation}
and the known functions $G_j(k)$, $j=1,2,3$, are defined in terms of $f$ as
follows:
\begin{mathletters}
\label{2.13}
\begin{eqnarray}
G_1(k)=\frac{1}{\Delta(k)\a(-k)}\{&&[E(k)+A(-ik)E(ik)][F(-ik)-A(-ik)F(ik)]\\
  &&+[1-A(-ik)][A(-ik)E(ik)F(k)+E(k)F(-k)]\}\;. \nonumber
\end{eqnarray}
\begin{equation}
G_2(k)=\frac{F(k)+F(ik)}{\a(ik)}\;,
\end{equation}
\begin{equation}
G_3(k)=\frac{1}{\Delta(k)}\{[1-A(k)][F(-ik)-A(-ik)F(ik)]+[1-A(-ik)]
  [F(k)-A(k)F(-k)]\}\;.
\end{equation}
\end{mathletters}
Indeed, Eq.~(\ref{2.11}b) is Eq.~(\ref{2.8}).  Eliminating $\phi(-k)$
from Eqs.~(\ref{2.9}), (\ref{2.10}), we find
\begin{eqnarray}
\label{2.14}
\a(-k)\phi(-ik)&&+E(-ik)\psi(-)-A(-ik)[\a(-k)\phi(ik)+E(ik)\psi(+)]\nonumber\\
&&=F(-ik)+F(-k)-A(-ik)[F(ik)+F(-k)]\;.
\end{eqnarray}
Replacing in this equation $\phi(ik)$ by Eq.~(\ref{2.8}) and comparing with
Eq.~(\ref{2.7}) we find Eq.~(\ref{2.11}c).  Replacing $\psi(+)$ in terms of
$\psi(-)$ in Eq.~(\ref{2.14}), using Eq.~(\ref{2.11}c), we find
Eq.~(\ref{2.11}a).

Eq.~(\ref{2.1}) expresses $q(x,y)$ in terms of $\rho_j(k)$, and
Eqs.~(\ref{2.3}), (\ref{2.11}) express $\rho_j(k)$ in terms of the {\em
unknown} functions $\phi(ik)$, $\psi(-)$, and the known functions $G_j(k)$. 
The known functions give rise to the contribution
\begin{eqnarray}
\label{2.15}
G(x,y)=\frac{1}{2\pi}\int_{\ell_1}e(k,z,\z)[\a(k)G_1(k)-F(-ik)]\frac{dk}{k}
+\frac{1}{2\pi}&&\int_{\ell_2}e(k,z,\z)[\a(-ik)G_2(k)-F(k)]\frac{dk}{k}
  \nonumber\\
&&+\frac{1}{2\pi}\int_{\ell_3}e(k,z,\z)E(k)G_3(k) \frac{dk}{k}\;.
\end{eqnarray}
In what follows we will show that, by using appropriate contour rotations,
the integrals involving the functions $\phi(ik)$, $\psi(-)$ can be evaluated
in terms of residues.  Furthermore, these residues can be computed in terms
of the functions $G_j(k)$.  For the justification of these rotations we use
the following facts (see Figure~3).

\smallskip\noindent
$\bullet$ $e(k,z,\z)$, $e(k,z,\z)E(k)$, $e(k,z,\z)E(-ik)$, are bounded for
$0<\arg k<\case{\pi}{2}$,
$\case{\pi}{2}<\arg k<\case{5\pi}{4}$, $\case{5\pi}{4}<\arg
k<2\pi$, respectively.

\smallskip\noindent
$\bullet$ $E(-k)E(ik)$ and $\psi(-)$ are bounded for $-\case{\pi}{4}<\arg
k<\case{3\pi}{4}$, while $E(k)E(-ik)\psi(-)$ is bounded for
$\case{3\pi}{4}<\arg k<\case{7\pi}{4}$.

\smallskip\noindent
$\bullet$ $\Delta(k)\sim E(k)$, $k\to0$ and $k\to\infty$, in
$-\case{\pi}{4}<\arg k<\case{3\pi}{4}$; 
$\Delta(k)\sim E(ik)$, $k\to0$ and $k\to\infty$, in $\case{3\pi}{4}<\arg
k<\case{7\pi}{4}$.

\smallskip\noindent
Indeed, since $x\geq0$ and $y\geq0$, $e(k,z,\z)$ is bounded both at $k=0$ and
$k=\infty$ in the first quadrant of the complex $k$-plane.  Since
$-\case{\pi}{2}<\arg(z-z_3)<-\case{\pi}{4}$, it follows that if
$\case{\pi}{2}<\arg k<\case{5\pi}{4}$ then $0<\arg k(z-z_3)<\pi$. 
Hence
$\exp[ik(z-z_3)-i\bk(\z-\z_3)]$ is bounded both at $k=0$ and $k=\infty$;
using $z_3=iL$, this exponential equals $e(k,z,\z)E(k)$.  Similar
considerations apply to $e(k,z,\z)E(-ik)$.

$\Delta(k)=E(ik)[E(k)E(-ik)+A(k)A(-ik)]$.  If $-\case{\pi}{4}<\arg
k<\case{3\pi}{4}$,
$E(k)E(-ik)$ is exponentially large at $k=0$ and $k=\infty$, and
$\Delta(k)\sim E(k)$.  Similarly, if $\case{3\pi}{4}<\arg
k<\case{7\pi}{4}$,
$E(k)E(-ik)$ is exponentially small, and $\Delta(k)\sim
E(ik)A(k)A(-ik)\sim E(ik)$.

$\psi(-)$ involves $-(ke^{-i\pi/4}+\bk e^{i\pi/4})$, thus it is
bounded for $-\case{\pi}{4}<\arg k<\case{3\pi}{4}$.  Similarly for
$E(k)E(-ik)\psi(-)$.

The contribution of the integral along $\ell_3$, due to the terms involving
$\psi(-)$ (see Eq.~(\ref{2.11}c)), gives rise to two integrals: one involving
$e(k,z,\z)A(k)E(-ik)E(k)\psi(-)/k\Delta(k)$, and one involving
$e(k,z,\z)A(-ik)E(k)^2/k\Delta(k)$.  The first
integral is bounded in $\case{5\pi}{4}<\arg k<2\pi$, while the second
integral is bounded in $\case{\pi}{2}<\arg k<\case{5\pi}{4}$.  Indeed,
the integrand of the first integral is dominated by
\[
[e(k,z,\z)E(-ik)][E(k)E(-ik)\psi(-)]\;,\quad\case{5\pi}{4}<\arg
k<\case{7\pi}{4}\;;\qquad [e(k,z,\z)E(-ik)][\psi(-)]\;,
\quad\case{7\pi}{4}<\arg k<2\pi\;,
\]
and each of the brackets is bounded.  Similarly, the integrand of the second
integral is dominated by
\[
[e(k,z,\z)E(k)][\psi(-)]\;,
\quad\case{\pi}{2}<\arg k<\case{3\pi}{4}\;;\qquad
[e(k,z,\z)E(k)][E(k)E(-ik)\psi(-)]\;,\quad\case{3\pi}{4}<\arg
k<\case{5\pi}{4}\;,
\]
and each of the brackets is bounded.

Hence, the integral along $\ell_3$, due to the terms involving $\psi(-)$,
equals an integral along $\ell_1$ involving
$e(k,z,\z)A(k)E(-ik)E(k)\psi(-)/k\Delta(k)$, an integral along $\ell_2$
involving
$e(k,z,\z)A(-ik)E(k)^2/k\Delta(k)$, and a contribution due to residues which
will be computed below (see Eqs.~(\ref{2.19}b), (\ref{2.19}c)).  Combining
these integrals with the integrals due to
$\phi(-ik)$ and to $\phi(k)$ (see Eqs.~(\ref{2.11}a) and (\ref{2.11}b)), we
find
\begin{eqnarray}
&&J_1(x,y)=\frac{1}{2\pi}\int\limits_{-\ell_2\cup\ell_1}
  e(k,z,\z)[\case{i}{2}q(0,0)+\a(k)A(-ik)\phi(ik)]\frac{dk}{k}\;,\\
\label{2.16} 
&&J_2(x,y)=\frac{1}{2\pi}\int\limits_{-\ell_2\cup\ell_1} 
\frac{e(k,z,\z)}{\Delta(k)}A(k)A(-ik)^2E(k)E(ik)\psi(-)\frac{dk}{k}\;.\label{2.17}
\end{eqnarray}
For $k$ in the first quadrant of the complex $k$-plane, $E(k)/\Delta(k)$ is
dominated by $1$, and each of the terms $e(k,z,\z)$, $\phi(ik)$, $E(ik)$,
$\psi(-)$ is bounded.  Thus, both $J_1$ and $J_2$ can be computed in terms of
residues.

The definition of $A(k)$ implies
\[
A(-ik)=-\frac{(k+\l_1)(k+\l_2)}{(k-\l_1)(k-\l_2)},\quad
  A(k)=-\frac{(k+\l_1)(k-\l_2)}{(k-\l_1)(k+\l_2)},\qquad
\l_1=\frac{\g}{4}(1+i),\quad \l_2=\frac{\g}{4}(-1+i)\;,
\]
so the poles of $A(-ik)$ and $A(k)$ occur at $\l_1$, $\l_2$, and at $\l_1$,
$-\l_2$, respectively.  Similarly, the poles of $A(ik)$ and $A(-k)$ occur at
$-\l_1$, $-\l_2$, and $-\l_1$, $\l_2$, respectively.  Using these facts it
follows that
\begin{equation}
\label{2.18}
q(x,y)=G(x,y)+\sum_{j=1}^3R_j(x,y)+P(x,y)\;,
\end{equation}
where $G(x,y)$ is defined by Eq.~(\ref{2.15}), $P(x,y)$ is the contribution to
$J_1$ and $J_2$ due to the poles of $\a(k)A(-ik)$ and of $A(k)A(-ik)^2$, and
the $R_j$ are defined as follows:
\begin{mathletters}
\label{2.19}
\begin{eqnarray}
&&R_1=i\sum_j\frac{e(\kj,z,\z)A(\kj)A(-i\kj)^2E(\kj)E(i\kj)}
  {\kj\Delta'(\kj)}\psi(\kj)\;,\\
&&R_2=-i\sum_j\frac{e(\lj,z,\z)A(-\lj)E(\lj)^2}{\lj\Delta'(\lj)}\psi(\lj)
  +2\frac{e(\l_2,z,\z)}{\Delta(\l_2)}E(\l_2)^2\psi(\l_2)\;,\\
&&R_3=-i\sum_j\frac{e(\mj,z,\z)A(\mj)E(-i\mj)E(\mj)}{\mj\Delta'(\mj)}\psi(\mj)
  -2\frac{e(-\l_2,z,\z)}{\Delta(-\l_2)}E(-\l_2)E(i\l_2)\psi(-\l_2)\;,
\end{eqnarray}
\end{mathletters}
$\psi(k)$, $\Delta'(k)$ denote
\begin{equation}
\psi(k)=\psi(-ke^{-i\pi/4})\;,\qquad\Delta'(k)=\frac{d\Delta(k)}{dk}\;,
\end{equation}
and  $\kj$, $\lj$, $\mj$ denote the zeros of $\Delta(k)$ in $0<\arg
k<\case{\pi}{2}$, $\case{\pi}{4}<\arg k<\case{5\pi}{4}$, $\case{5\pi}{4}<\arg
k<2\pi$, respectively.
Multiplying Eq.~(\ref{2.11}c) by
$\Delta(k)$ and evaluating the resulting expression at $k_j=\{\kj,\lj,\mj\}$,
we find
\begin{equation}
\label{2.21}
\psi(k_j)=-\frac{[1-A(k_j)][F(-ik_j)-A(-ik_j)F(ik_j)]
   +[1-A(-ik_j)][F(k_j)-A(k_j)F(-k_j)]}{\delta(k_j)}\;,
   \quad\delta(k_j)\neq0\;,
\end{equation}
where
\begin{equation}
\label{2.22}
\delta(k)=A(-ik)E(k)+A(k)E(-ik)\;,
\end{equation}
Noting that $\a(k)=(k+\l_1)(k-\l_2)/2k$, $\a(ik)=-(k-\l_1)(k-\l_2)/2ik$, and
evaluating Eq.~(\ref{2.8}) at $k=\l_2$, we find $\psi(\l_2)$.  Similarly,
evaluating Eq.~(\ref{2.9}) at $k=-\l_2$ we find $\psi(-\l_2)$:
\begin{equation}
\label{2.23}
\psi(\l_2)=E(-\l_2)[F(\l_2)+F(i\l_2)]\;,\qquad
\psi(-\l_2)=E(-i\l_2)[F(\l_2)+F(i\l_2)]\;.
\end{equation}

The term $P(x,y)$ arises from $\a(k)A(-ik)$ in $J_1$, and
$A(k)A(-ik)^2/\Delta(k)$ [$\Delta(k)\neq0$] in $J_2$, each of which has a
simple pole at $k=\l_1$.  Evaluation of the pertaining residues yields
\[
P(x,y)=2e(\l_1,z,\z)[\a(\l_1)\phi(i\l_1)+E(\l_1)\psi(\l_1)]\;.
\]
Evaluating Eq.~(\ref{2.8}) at $k=\l_1$, we find
\[
\a(\l_1)\phi(i\l_1)+E(\l_1)\psi(\l_1)=F(\l_1)+F(i\l_1)\;.
\]
Thus,
\begin{equation}
\label{2.24}
P(x,y)=2e(\l_1,z,\z)[F(\l_1)+F(i\l_1)]\;.
\end{equation}

\medskip\noindent
\und{In summary}: {\em Assume that $\delta(k_j)\neq0$, where $k_j$ is a zero
of $\Delta(k)$, and $\delta(k)$, $\Delta(k)$ are defined by
Eqs.~(\ref{2.12}b), (\ref{2.22}), respectively. Then
$q(x,y)$ is given by Eq.~(\ref{2.18}), where
$G(x,y)$ is defined by Eq.~(\ref{2.15}), $P(x,y)$ is defined by
Eq.~(\ref{2.24}), and $R_j(x,y)$, $j=1,2,3$, are defined by Eqs.~(\ref{2.19}),
with $\psi(k_j)$, $\psi(\l_2)$, $\psi(-\l_2)$ defined by Eqs.~(\ref{2.21}),
(\ref{2.23}).}

\section{The physical problem}
The physical problem corresponds to the homogeneous BCs, {\it i.e.}, $f=0$.
In this case Eq.~(\ref{2.18}) yields $q(x,y)=0$.  Thus, we only need to
consider the assumption that $\delta(k_n)\neq0$.  If this assumption is
violated then the equations $\Delta(k_n)=0$ and $\delta(k_n)=0$ can be
rewritten in the form 
\begin{mathletters}
\label{06}
\begin{eqnarray}
A(ik_n)^2E(-ik_n)^2 &=& 1\;, \\ 
A(-k_n)^2E(k_n)^2 &=& 1\;.  
\end{eqnarray}
\end{mathletters}
Eqs.~(\ref{06}) do not have a solution for generic values of $\g$. 
Indeed, consider first the limit of infinite back reaction rate, $\g
L\to\infty$.  Inspection of Eqs.~(\ref{06}) in this limit yields the
asymptotic solution $k_{\infty}=\pm\l_1$, $\pm\l_2$.  If there exists a steady
state other than $q(x,y)=0$, then it would also exist for $\g L$
large but finite.  We therefore seek solutions of~(\ref{06}) of the form
$k=k_{\infty}+\epsilon$.  Such solutions do not exist: Using $k_\infty=\l_1$,
Eq.~(\ref{06}a) yields
$\epsilon=\case{1}{4L}(-1+i)$ --- to first order in $\epsilon$ ---
while Eq.~(\ref{06}b) yields the contradictory result
$\epsilon=\case{1}{4L}(+1-i)$.  The other values of $k_\infty$ lead to
similar contradictory results.  Thus, the only solution to the physical
problem is
$q(x,y)=0$, which corresponds to the trivial case of the vacuum; when no
particles are left in the system.

Finally, consider the relaxation of the system into the absorbing empty
state.  Instead of Eq.~(\ref{E.eq}), we need to study
\begin{equation}
E_{xx}+E_{yy}+\g(-E_x+E_y)=E_{\tau}\;,
\end{equation}
where $\tau=Dt$ is a rescaled time parameter.  
We turn this into an eigenvalue problem, by writing
$E(x,y,t)=1-e^{-\sigma\tau}e^{-{\g\over2}(y-x)}q_{\sigma}(x,y)$.  This
results in an equation for $q_{\sigma}$ identical to Eq.~(\ref{helmholtz}),
valid over the same domain, but with
$4\b^2=\case{1}{2}\g^2-\sigma$.  The BCs for this equation are identical
to~(\ref{bcs}).  We have already seen that the problem admits no zero
eigenvalue: $q_0(x,y)=0$.  The analysis for $\sigma>0$ proceeds along the
same lines.  Once again, the critical issue is whether there exist solutions
of Eqs.~(\ref{06}).  This time the asymptotic solution for $\g
L\to\infty$ is $k_{\infty}=\case{\g}{4}[\pm1\pm
i(1-\case{4\sigma}{\g})^{1/2}]$, 
$\case{\g}{4}[\pm(1-\case{4\sigma}{\g})^{1/2}\pm i]$.  A perturbation
analysis shows that solutions exist for finite $\g L\ll1$,   
provided that
$\sigma\sim2\g^2e^{-\g L/2}$.  The relaxation time to the
empty state is therefore $(D\sigma)^{-1}=(D/2v^2)e^{vL/2D}$.

It is instructive to compare our analysis of $A+A\rightleftharpoons A$ to the
mean-field result.  The reaction-diffusion equation for the
steady state of the process, in a segment demarcated by traps,
is
\begin{equation}
\label{mf}
D\rho_{xx}+k_1\rho-k_2\rho^2=0\;,\qquad -L/2\leq x\leq L/2\;,
\end{equation}
where $\rho(x)$ is the local particle density, $k_1$ is the rate of the back
reaction $A\to A+A$, $k_2$ is the rate of $A+A\to A$, and the traps impose
the BCs $\rho(\pm L/2)=0$. This equation predicts a transition from an empty
state ($\rho=0$) to an active state ($\rho>0$), when $k_1$ exceeds a certain
critical value~\cite{vonhaeften97,remark}.  Our exact
analysis shows that in the actual system of one-dimensional coalescence the
noise destroys the transition and the only existing steady state is the empty
state.  The nontrivial steady state of the mean-field case is echoed
in the exponentially large relaxation time found for large back reaction
rates.  Although the lack of a transition cannot be established
from numerical simulations, especially in view of the long relaxation
times for $\g L$ large, previous work had suggested that a transition does
not take place~\cite{ner99}.

\acknowledgments
We thank Prof.~L.~Glasser and Prof.~D.~Kessler for useful discussions, and
Prof.~A.~C.~Newell for important suggestions.  We gratefully acknowledge the
NSF (D.b.-A.) and the EPRSC (A.S.F.) for support of this work.


\begin{figure}
\centerline{\epsfxsize=5cm \epsfbox{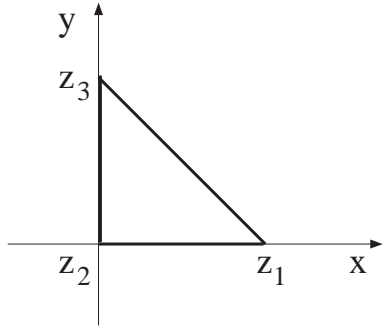}}
\caption{Domain of the modified Helmholtz equation, Eq.~(\ref{helmholtz}).}
\end{figure}

\begin{figure}
\centerline{\epsfxsize=5cm \epsfbox{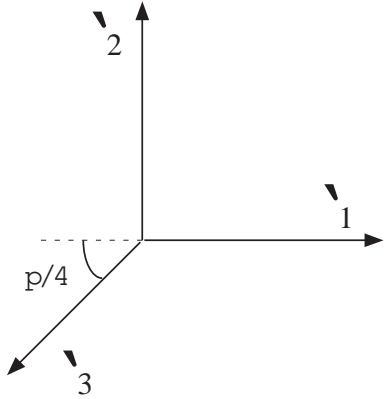}}
\caption{The rays $\ell_j$, in the complex plane, along which $q(x,y)$ is
computed (Eq.~(\ref{2.1})).}
\end{figure}

\begin{figure}
\centerline{\epsfxsize=16cm \epsfbox{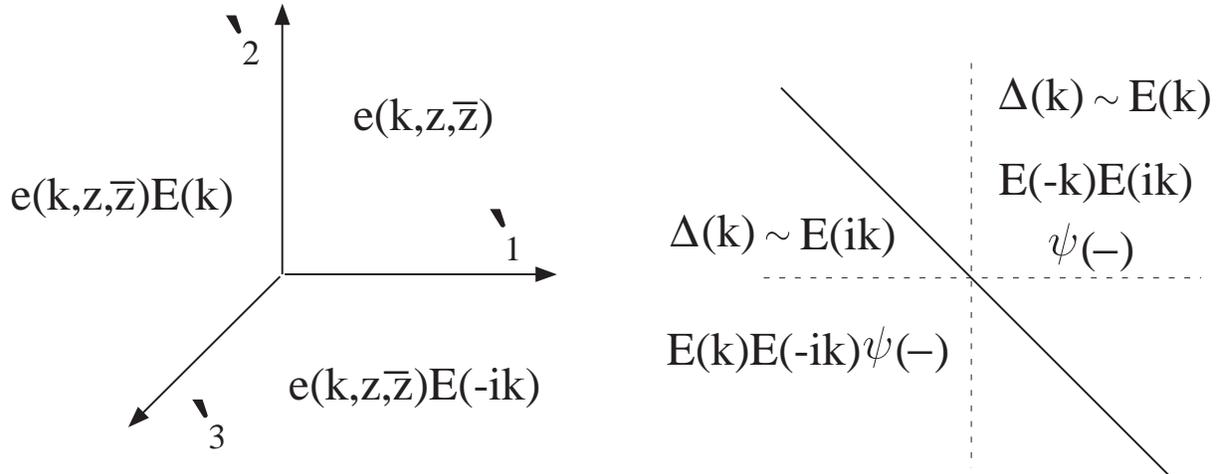}}
\caption{Regions where $e(k,z,\z)$, $e(k,z,\z)E(k)$, $e(k,z,\z)E(-ik)$,
$\psi(-)$,
$E(k)E(-ik)\psi(-)$ are bounded, and dominant behavior of $\Delta(k)$.}
\end{figure}


\begin{thebibliography}{99}
\bibitem{remark} This transition is familiar to nuclear
physicists: the three-dimensional version of Eq.~(\ref{mf}) describes the
cascading fission of nuclei in a lump of radioactive material.  If $k_1$ is
held constant, the transition takes place when $L$ is sufficiently large,
giving rise to the concept of ``critical mass".
\end{thebibliography}
\end{document}